\documentclass[slac_one]{revtex4}
\usepackage{graphicx,amssymb,amsmath}
\usepackage{fancyhdr}
\usepackage{latexsym}
\usepackage[dvips]{color}
\newcommand{\EPEM}{\mbox{$e^+e^{-}$}}
\newcommand{\EMEM}{\mbox{$e^-e^-$}}
\newcommand{\GG}{\mbox{$\gamma\gamma$}}
\newcommand{\GE}{\mbox{$\gamma e$}}

\newcommand{\LGG}{\mbox{$L_{\gamma\gamma}$}}

\newcommand{\LEPEM}{\mbox{$L_{e^+e^-}$}}

\newcommand{\MKM}{\mbox{$\mu$m}}

\newcommand{\be}{\begin{equation}}
\newcommand{\ee}{\end{equation}}
\newcommand{\bc}{\begin{center}}
\newcommand{\ec}{\end{center}}
\newcommand{\bi}{\begin{itemize}}
\newcommand{\ei}{\end{itemize}}
\newcommand{\ben}{\begin{enumerate}}
\newcommand{\een}{\end{enumerate}}
\begin{document}
\title{{\small{2005 ALCPG \& ILC Workshops - Snowmass,
U.S.A.}}\\ %% Please keep this conference title here
\vspace{12pt} Physics Options at the ILC. \\ GG6\footnote{GG6
 Snowmass home page:
http://alcpg2005.colorado.edu:8080/alcpg2005/program/accelerator/GG6/agenda.  \\
Participants: T.~Omori, convener (KEK), B.~Parker, convener (BNL),
P.~Bambade (LAL), G.~Gronberg (LLNL), C.~Heusch (UCSC), S.~Kanemura
(Osaka), Yu.~Kolomensky (LBNL), K.~Kubo (KEK), D.~J.~Miller (UCL),
K.~M\"{o}nig (DESY), S.~Mtingwa (N.Carolina U.), D.~Scott (Daresbury),
A.~Seryi (SLAC), A.~Wolski (LBNL).} Summary at Snowmass2005}


\author{V.~I~.Telnov}
\affiliation{Budker Institute of Nuclear Physics, 630090 Novosibirsk, Russia}
\begin{abstract}
  At Snowmass2005 the Global Group 6 (Physics Options) considered the
  requirements and configurational issues related to possible
  alternatives to the baseline \EPEM\ collisions, including \GG, \GE,
  \EMEM, GigaZ and fixed target experiments, and identified the potential
  performance parameters.

\end{abstract}

\maketitle

\thispagestyle{fancy}
\section{PHOTON COLLIDER}

\subsection{Introduction}

\begin{figure}[htb]
%\vspace{-1.6cm}
\centering
\includegraphics[width=10cm]{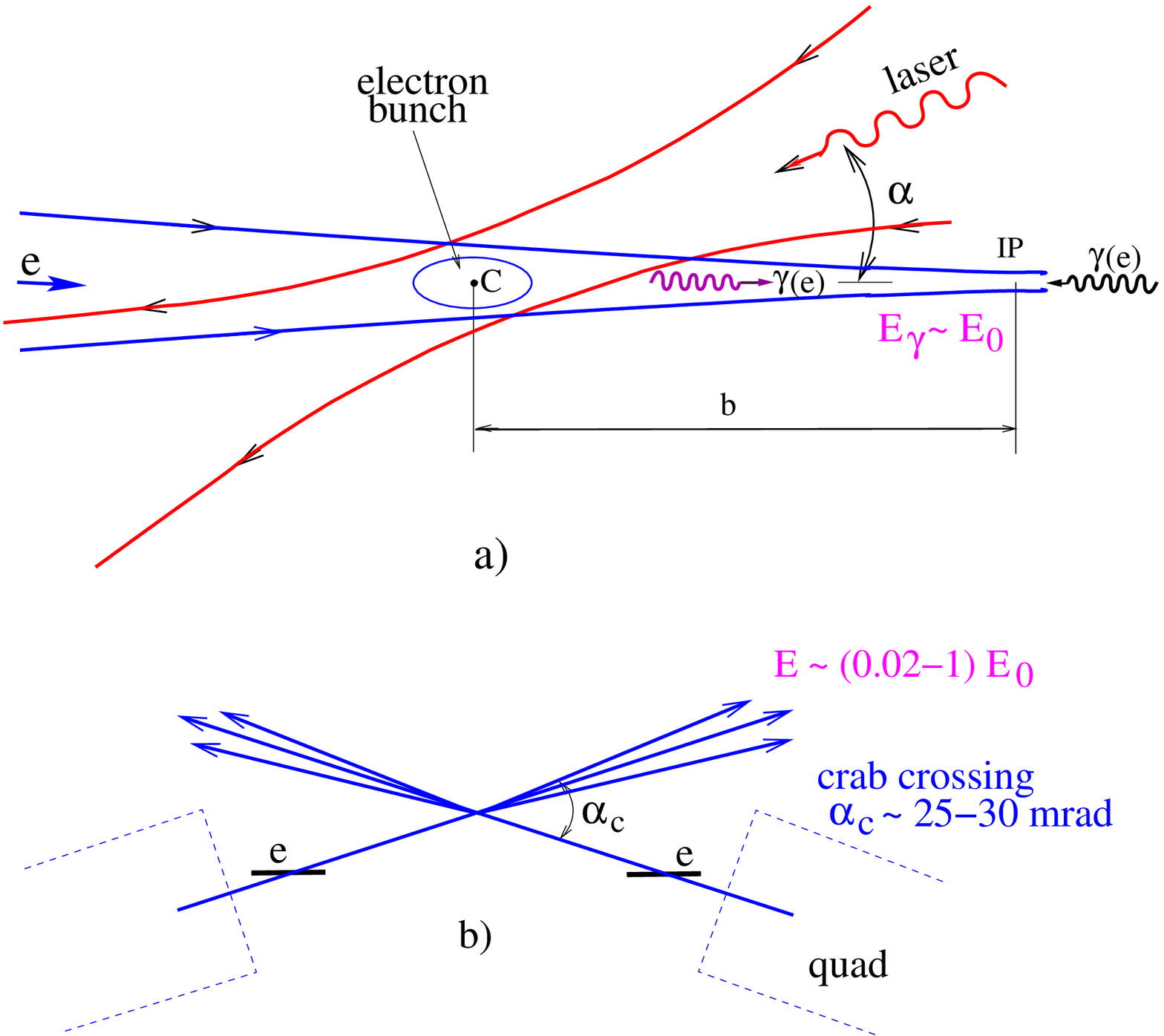}
%\vspace{-1.cm}
\caption{Scheme of \GG,\GE\ collider}
\label{scheme}
\end{figure}

At the photon collider, \GG\ and \GE\ collisions are obtained by
"conversion" of electrons into high-energy photons using Compton
scattering of laser light at the distance $b \sim \gamma \sigma_y
\sim$ 1--2 mm from the interaction point (IP), Fig.~\ref{scheme}. For
the energy $2E_0=500$ GeV, the optimum wavelength (determined by the
threshold for  \EPEM\ pair creation at the conversion region) is
about 1 \MKM, which coincides with that of the most powerful lasers
available today. The maximum energy of photons is about 80\% of the electron
beam energy, so the maximum invariant mass of the
\GG\ system is about 80\% of the c.m.s. beam energy; for \GE\
collisions it is  90\%.  Lower invariant masses are obtained by decreasing
the electron beam energy. For example, to study  the Higgs boson of
120 GeV mass, one needs $2E_0$=200 GeV. The most comprehensive description
of the photon collider available at present that is the TESLA TDR
\cite{TESLATDR}; almost all considerations done for TESLA are valid
for the ILC as well.

At the nominal ILC parameters, the expected \GG\ luminosity in the high-energy peak of the luminosity spectrum $\LGG\sim 0.17\, \LEPEM$~\cite{Telnov_snow05_1}.
However, the \GG\ luminosity at the ILC is not restricted by collision
effects and can be increased by a factor of 2--3 by reducing emittances
in damping rings (which is not easy but possible), reaching $\LGG\sim$
(0.3--0.5) \LEPEM.  Typical cross sections of many interesting processes in
\GG\ collisions (charged pairs, Higgs, etc) are higher than those in
\EPEM\ collisions by about one order of magnitude, so the number of events in
\GG\ collisions will be larger than in \EPEM\ even for the nominal
beam parameters foreseen for \EPEM\ collisions.

The physics program of the photon collider is very rich and complements in
an essential way the physics in \EPEM\ collisions,  independently of the
physics scenario. In \GG, \GE\ collisions, compared to \EPEM,
\bi
\item the energy is smaller only by 10--20\%; \\[-8mm]
\item the number of interesting events is similar or even larger; \\[-8mm]
\item access to higher particle masses (H,A in \GG, SUSY in
  \GE);  \\[-8mm]
\item higher precision for some phenomena; \\[-8mm]
\item different types of  reactions;  \\[-8mm]
\item highly polarized photons.  \ei 
In summary, the physics reach of a \GG, \GE\ collider is not worse
than that of a \EPEM\ collider it is based on. The only advantage of
\EPEM\ collisions is the narrower
  luminosity spectrum, a feature that is of rather limited use.
  It is very important to note
  that the photon collider can be added to the linear \EPEM\ collider
  nearly for free (the laser system, modification of the IP and one of
  detector add less than 3--4\% of the ILC cost). The decrease of the
  \EPEM\ running time by 25--30\% is a neglegible price to pay for the
  opportunity to look for new phenomena  in other types of
  collisions.

\subsection{Special requirements for the ILC}

The photon collider presents several special requirements that should
be taken into account in the baseline ILC design:
\bi

\item For the removal of  disrupted beams, the crab-crossing angle at one of
  the interaction regions should be about 25 mrad (the exact number
  depends on the quad design); the quad fringe field should
  be small in the region of the outgoing low-energy beam;
  
\item The \GG\ luminosity is nearly proportional to the geometric
  \EMEM\ luminosity, so the product of the horizontal and vertical
  emittances should be as small as possible (this translates into
  requirements on the damping rings and beam transport lines);

\item The final-focus system should provide beam spot sizes at the
  interaction point as small as possible (compared to the \EPEM\ case, the
  horizontal $\beta$-function should be smaller by one order of
  magnitude);
\item The very wide disrupted beams should be transported to the beam dumps
  with acceptable losses. The beam dump should be able to withstand absorption of
  very narrow photon beam after the Compton scattering;
\item  The detector design should allow easy replacement of elements in the
forward region  ($<$100 mrad);
\item Space for the laser and laser beam lines  has to be reserved.
\ei

\subsection{Crossing angle}.

\begin{figure}[!htb]
\vspace{-1.cm}
\centering
\includegraphics[width=11cm]{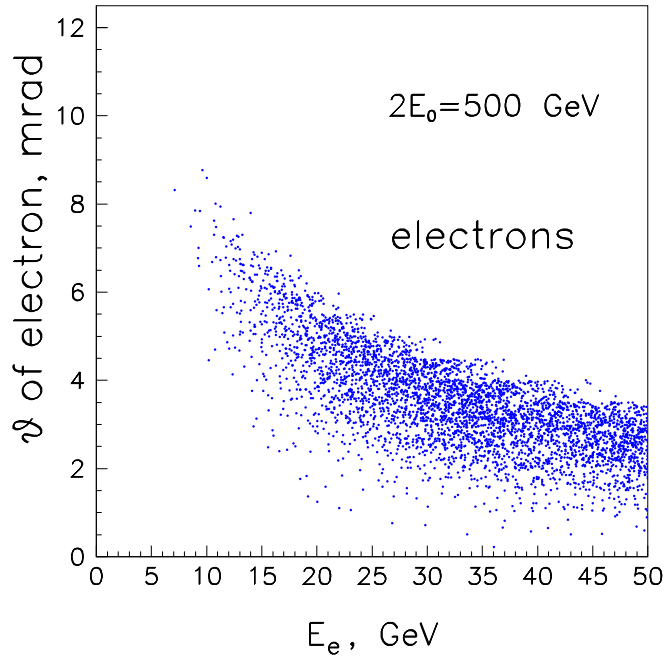}
\vspace{-1.cm}
\caption{Angles of disrupted electrons after Compton scattering and
  interaction with opposing electron beam; $N=2\times 10^{10}$,
  $\sigma_z=0.3$ mm.}
\label{e-angle}
\end{figure}

After passing the conversion and collision points, the electrons have energy
from about 5 GeV up to $E_0$ and the horizontal disruption angle up to
about 10 mrad, see Fig.\ref{e-angle} (due to limited statistics in
simulation, about $10^5$ macroparticles, the maximum angles should be multiplied by a
factor of 1.2~ \cite{TESLATDR}).
Above this angle, the total energy of particles is less than that in
the secondary  irremovable \EPEM\ background.

For removal of these disrupted beams one needs the crab-crossing angle
to be larger than the disruption angle plus the angular size of the final
quad, see Fig.~\ref{scheme}.  Thare is an additional requirement: the field
outside the quad (in the region of the disrupted beam) should be small
in order to add a small deflection angle for low-energy particles.

\begin{figure}[!htb]
%\vspace{-0.6cm}
\centering
\includegraphics[width=10cm]{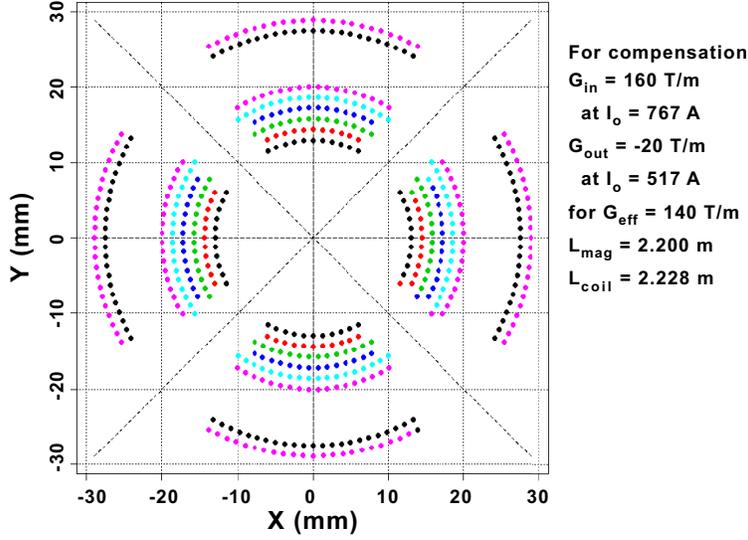}
\vspace{-0.6cm}
\caption{Design principle  of the superconducting quad (only coils are
  shown).  The radius of the quad with the cryostat is about 5 cm. The
  residual field outside the quad is negligibly small.}
\label{f-quad}
\end{figure}

Possible quad designs for the photon collider were considered at
Snowmass2005 by B.~Parker. In his initial design, the field outside the
superconducting quad was shielded by an active superconducting
screen (winding) around the disrupted beam~\cite{Parker}. This
solution is possible but rather difficult, and compensation is not
sufficient. At the workshop, B.~Parker produced another, much
more attractive quad design, which gives a minimum crossing angle, see
Fig.~\ref{f-quad}. In short, the quad consists of two quads of
different radiÛs, one inside another, with opposite field
directions. In this design the gradient on the axis is reduced only by
15\%, the field outside the quad is practically zero, so no additional
shielding is required.

\begin{figure}[!htb]
%\vspace{-0.6cm}
\centering
\includegraphics[width=7cm]{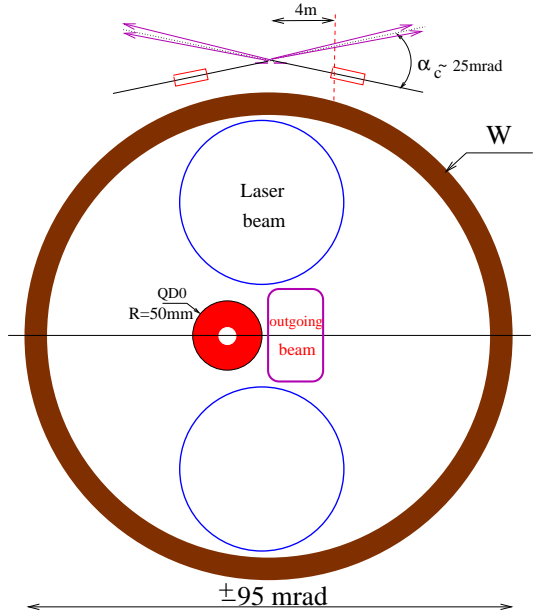}
%\vspace{-1.6cm}
\caption{Layout of the quad and electron and laser beams at the
  distance of 4 m from the interaction point (IP).}
\label{beams-quad}
\end{figure}

The radius of the quad, the cryostat taken into account, is $R=5$ cm. For the
horizontal disruption angle of 11 mrad (with about 20\% margin) the minimum
crab-crossing angle is 23--27 mrad for the distance of the quad from
the IP $L^*= $4.5--3.5 m, respectively. Obtaining the final numbers requires
some additional checks, but roughly it is 25 mrad. RElative positions of
the quad, the outgoing electron beam and the laser beam at the distance 4
m from the IP is shown in Fig. \ref{beams-quad}

Several important remarks concerning the crossing angle:

\bi

\item In considering the required crab-crossing angle we assumed that the
  bunch length $\sigma_z= 0.3$ mm. This gives $\alpha_c=25$ mrad. At
  present, the bunch length $\sigma_z= 0.15$ mm is also considered for
  \EPEM\ collisions. The disruption angle for low energy particles is
  proportional to $\sqrt{N/\sigma_z}$ and depends very weakly on
    transverse beam sizes. This means that for $\sigma_z=0.15$ mm the
    disruption angle is about 14--15 mrad and the required crab-crossing
    angle 30 mrad. So, shorting of the bunch length is not desirable
  for the photon collider.
\item In evaluating the disruption angles, we assume that the thickness of the
  laser ``target'' is equal to one scattering length for electrons
  with initial energy (65\% of electrons scatters). Further increase
  of the conversion probability leads to lowering the electron
  energies  due to multiple scattering and, correspondingly, to
  increasing disruption angles.
\item We assumed that the laser wavelength is $\lambda=1$ \MKM\ for
  all ILC energies. Due to pair creation in the conversion region,
  at the energy $2E_0>$700--800 GeV it is desirable that a 
  longer laser wavelength be used. The disruption angles in this case will be
  smaller.
\item For low-energy operation, say $E_0=100$ GeV, one can consider the
  possibility of using a shorter laser wavelength (doubled of
  triplet frequencies) in order to increase the energy of backscattered
  photons and improve the shape of the spectrum. However, this would lead to 
  increasing disruption angles. Fixing $\alpha_c = 25$ mrad, we close
  the possibility of shortening the laser wavelength. For physics this
  is acceptable, lower parameters $x=4E_0 \omega/mc^2$   even have
  some advantages for the Higgs study.

\begin{figure}[!htb]
%\vspace{-0.6cm}
\centering
\includegraphics[width=11cm]{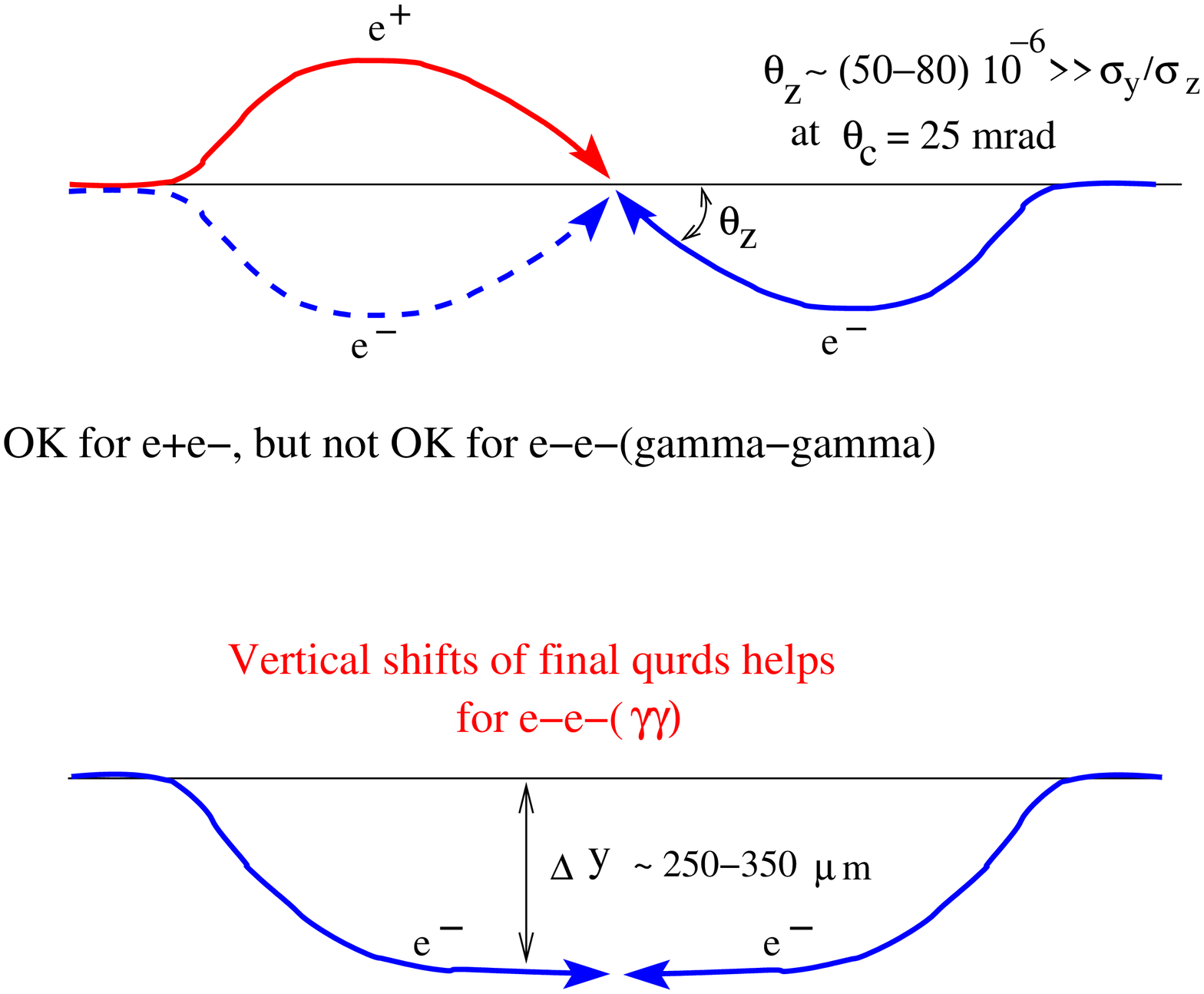}
%\vspace{-1.6cm}
\caption{Trajectories of electron (positron) in the presence of the
solenoid field and crab-crossing angle. At the lower figure, the
\EMEM\ collision angle is made zero using shifted quads.  }
\label{vert_angle}
\end{figure}

\item Due to the detector field \EMEM\ beams collide at a non-zero
  vertical collision angle which is several times larger than
  $\sigma_y/\sigma_z$, Fig.\ref{vert_angle}. This angle can be removed
  by dipole correction winding in quads~\cite{telnov_lcws05_1113}.
  Such a correction shifts the IP vertically by about 300 \MKM, which is
  acceptable.

\begin{figure}[!htb]
\vspace{-0.6cm}
\hspace{-0.0cm} \includegraphics[width=8cm]{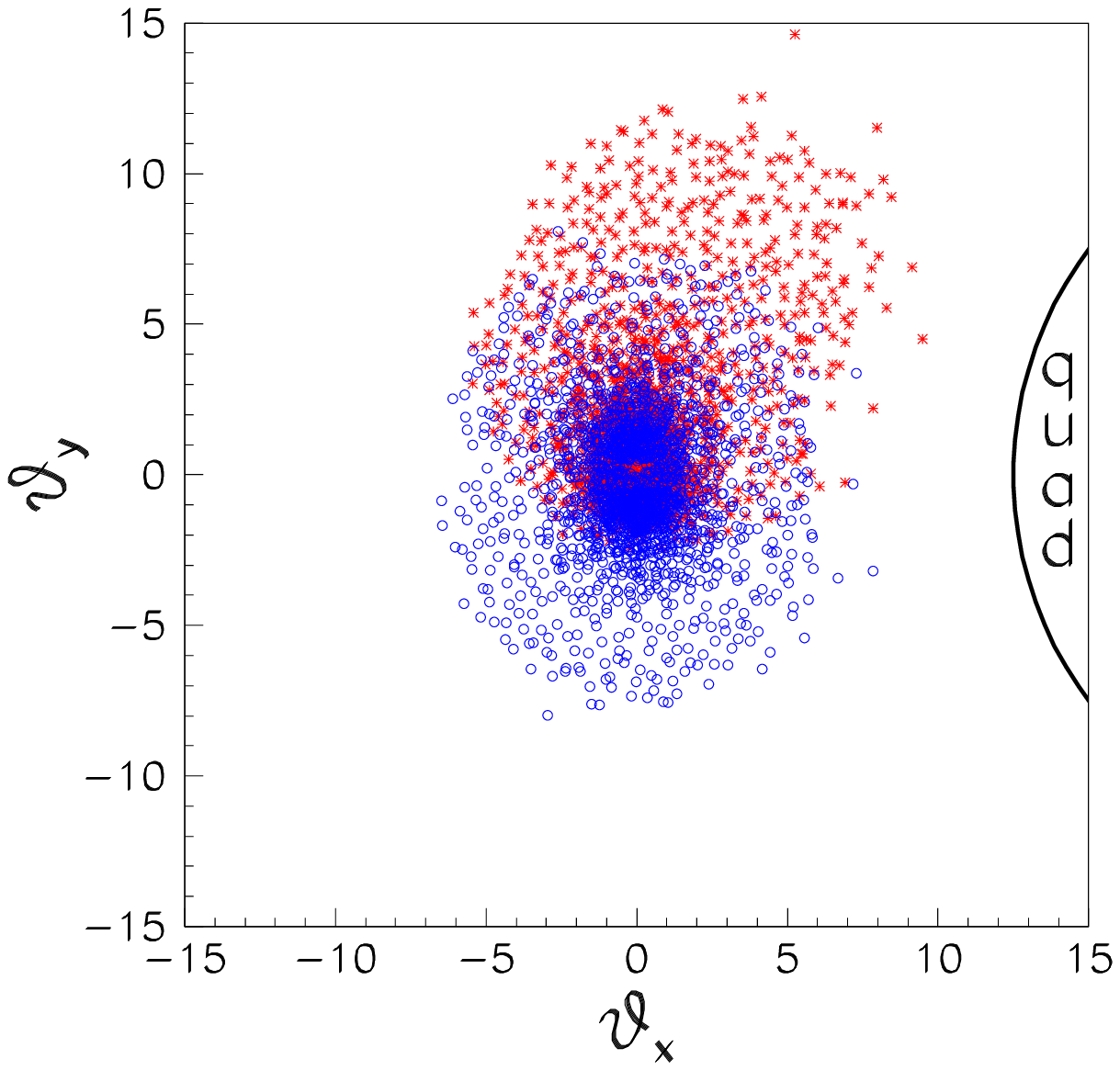}\hspace{-1cm}
\includegraphics[width=8cm]{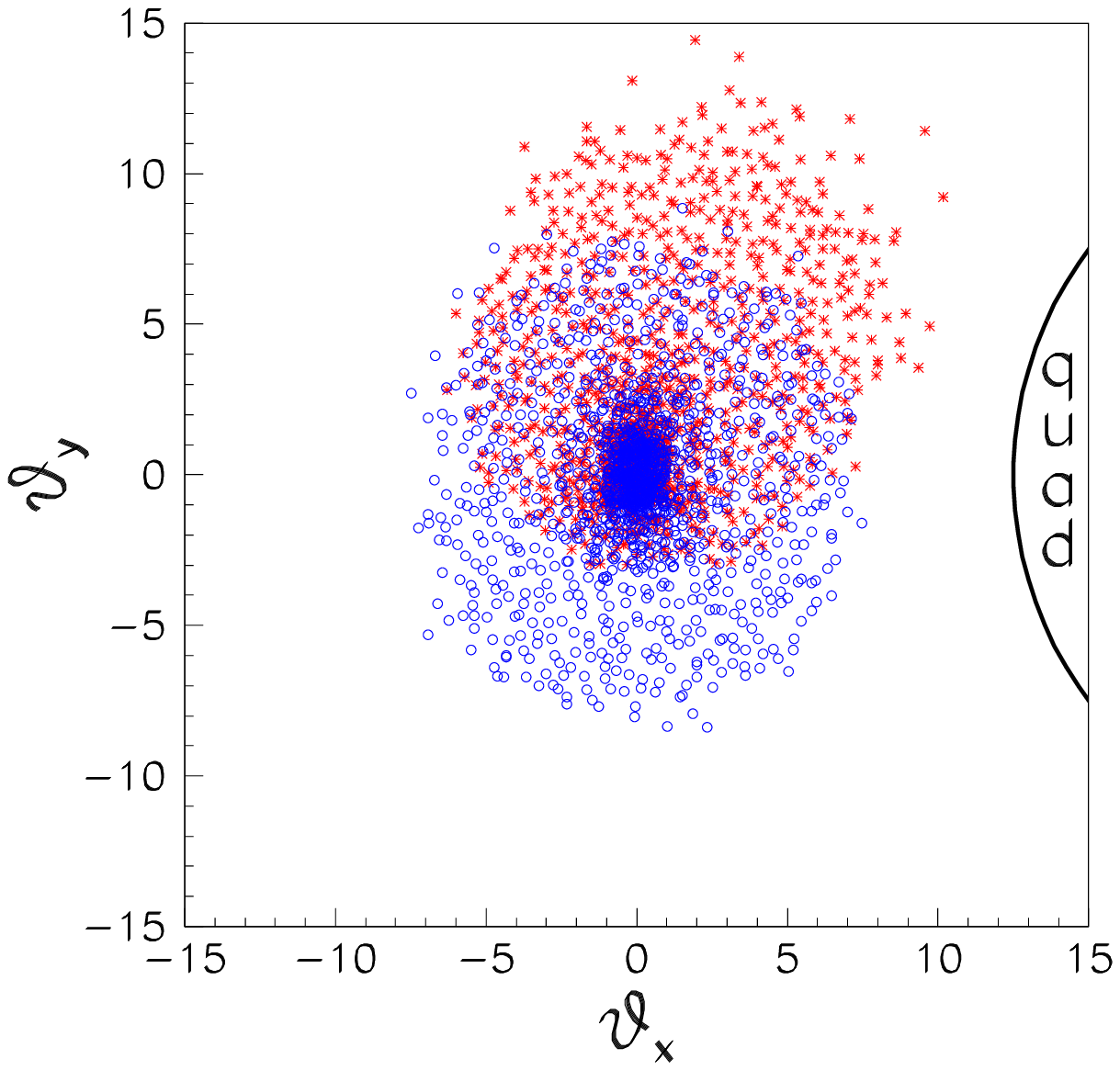}
\vspace{-0.9cm}
\caption{The shift of the disrupted beam due to the detector
  field. Blue (square)points: only beam-beam deflection, red (stars)
  points: the detector field of 4T is added. A crab-crossing angle of 25
  mrad and $2E_0=500$ GeV are assumed. Positions of particles are
  taken at the distance of 4 m from the IP, at the place where they
  pass the first quad. Left figure: $2E_0=200$ GeV, right: $2E_0=500$
  GeV. The total number of macroparticles in the beam (several
  collisions) is about 150000. Taking into account the tails, which can cause
  backgrounds, the ``effective'' beam sizes are larger by about 20\%.  }.
\label{thxthy}
\end{figure}

\begin{figure}[!htb]
\vspace{-1cm}
\centering
\includegraphics[width=10cm,height=9cm]{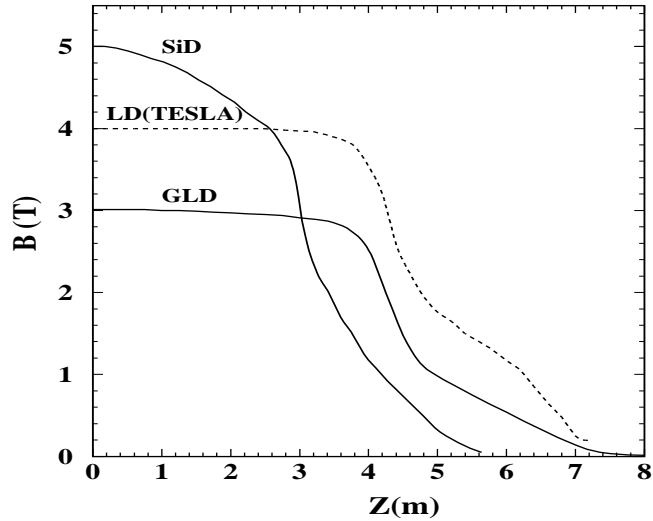}
\vspace{-1.5cm}
\caption{Magnetic field  $B(z,0,0)$ in LDC, SID and GLD detectors}
\label{bz}
\vspace{-0.cm}
\end{figure}

\item The solenoid field gives an additional deflection angle to the
  disrupted beam. The vertical deflection for lowest-energy particles
  (5 GeV) by the solenoid field is about 5 mrad, which adds to the 10--12 mrad
  acquired during the beam-beam collisions, so the total vertical angle
  is about 17 mrad. The solenoid field also leads to some horizontal
  displacement of the disrupted beam (due to vertical motion of
  particles) but it is smaller than the vertical shift of the beam.
  Figure \ref{thxthy} shows the influence of the detector field on the
  disrupted beam.  These figures correspond to head-on collisions at
  the photon collider with beam energies $2E_0= 200$ and 500 GeV. For
  beams with an initial mutual shift at the IP, the central core is
  shifted due to the instability of collisions but the maximum angles
  are practically do not change and decrease for large beams shifts.
\item Synchrotron radiation (SR) in the detector field leads to an
  increase of the vertical beam size. This effect was considered in
  Refs.~\cite{telnov_lcws05_1113, Telnov_snow05_1}.  Detector fields
  used in simulations are shown in Fig.~\ref{bz}.  Results of the
  simulation are presented in Table~\ref{tab2}; the statistical
  accuracy is about $\pm$0.5\%. We see that at the crossing angle of 25
  mrad the decrease of the \EPEM\ luminosity due to the SR is 5, 1.5
  and 2 \% for LDC, SiD and GLD detectors, respectively.  It is
  possible that possibly that by
  proper shaping of the field in the LDC the effect can be reduced further.
 \ei

\begin{table}[ht]
\bc
\caption{ Results on  $L(\alpha_c)/L(0)$.}  \setlength{\tabcolsep}{3.8mm}
 {  \EPEM\ collisions} \\[4mm]
\begin{tabular}{lllllll}
$\alpha_c$(mrad) & 0 & 20 & 25 & 30 & 35 & 40 \\ \hline
LDC(TESLA) & 1. &0.98 &0.95 &0.88 &0.83 & 0.76 \\
SID & 1. &0.995 &0.985 &0.98 &0.95 & 0.91 \\
GLD & 1. &0.995 &0.98 &0.97 &0.94  & 0.925 \\
\end{tabular} \\[0.5cm]
{ \GG\ collisions} \\[4mm]
\begin{tabular}{lllllll}
$\alpha_c$(mrad) & 0 & 20 & 25 & 30 & 35 & 40 \\ \hline
LDC(TESLA) & 1 &0.99 &0.96 &0.925 &0.86 & 0.79 \\
SID & 1 &0.99 &0.975 &0.955 &0.91 & 0.86 \\
GLD & 1 &0.995 &0.985 &0.98 &0.97 & 0.93
\end{tabular}\\
\ec
\label{tab2}
\end{table}

\begin{figure}[p]
\vspace{0.6cm} \hspace{0.5cm}
\includegraphics[width=12cm]{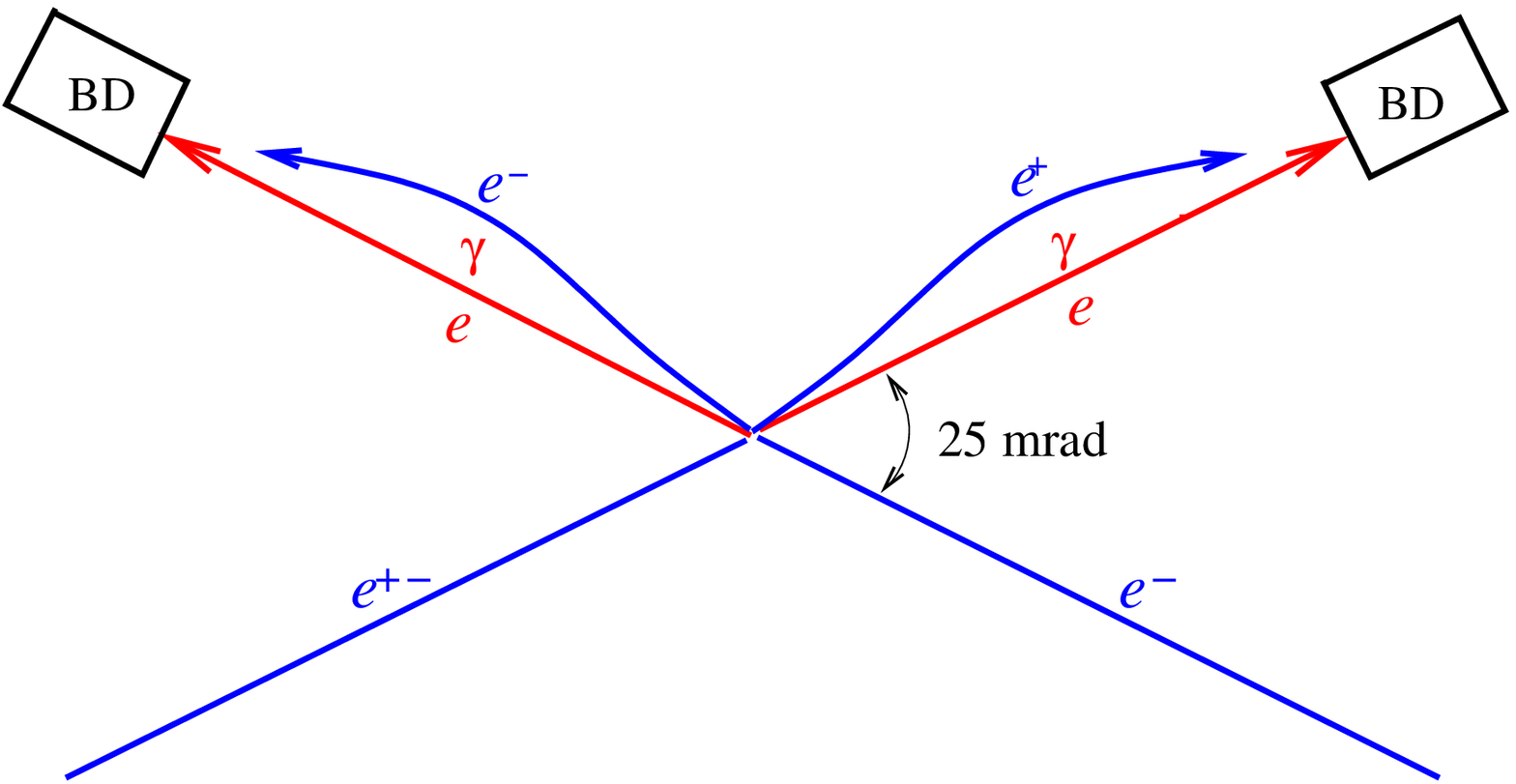} \vspace{1.5cm}
\hspace{0.5cm}
\includegraphics[width=12cm]{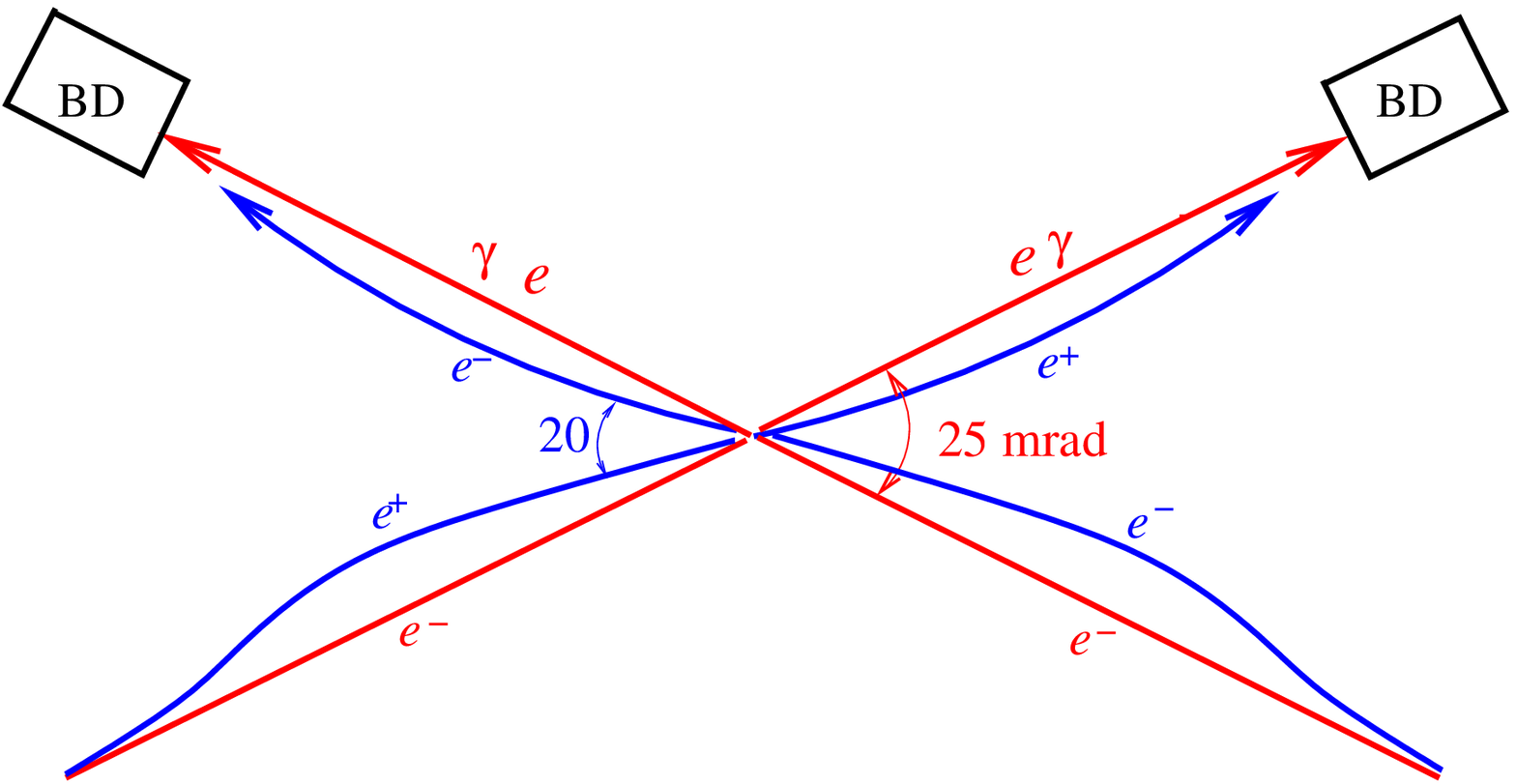} \vspace{1.5cm}
\hspace{0.5cm}
\includegraphics[width=12cm]{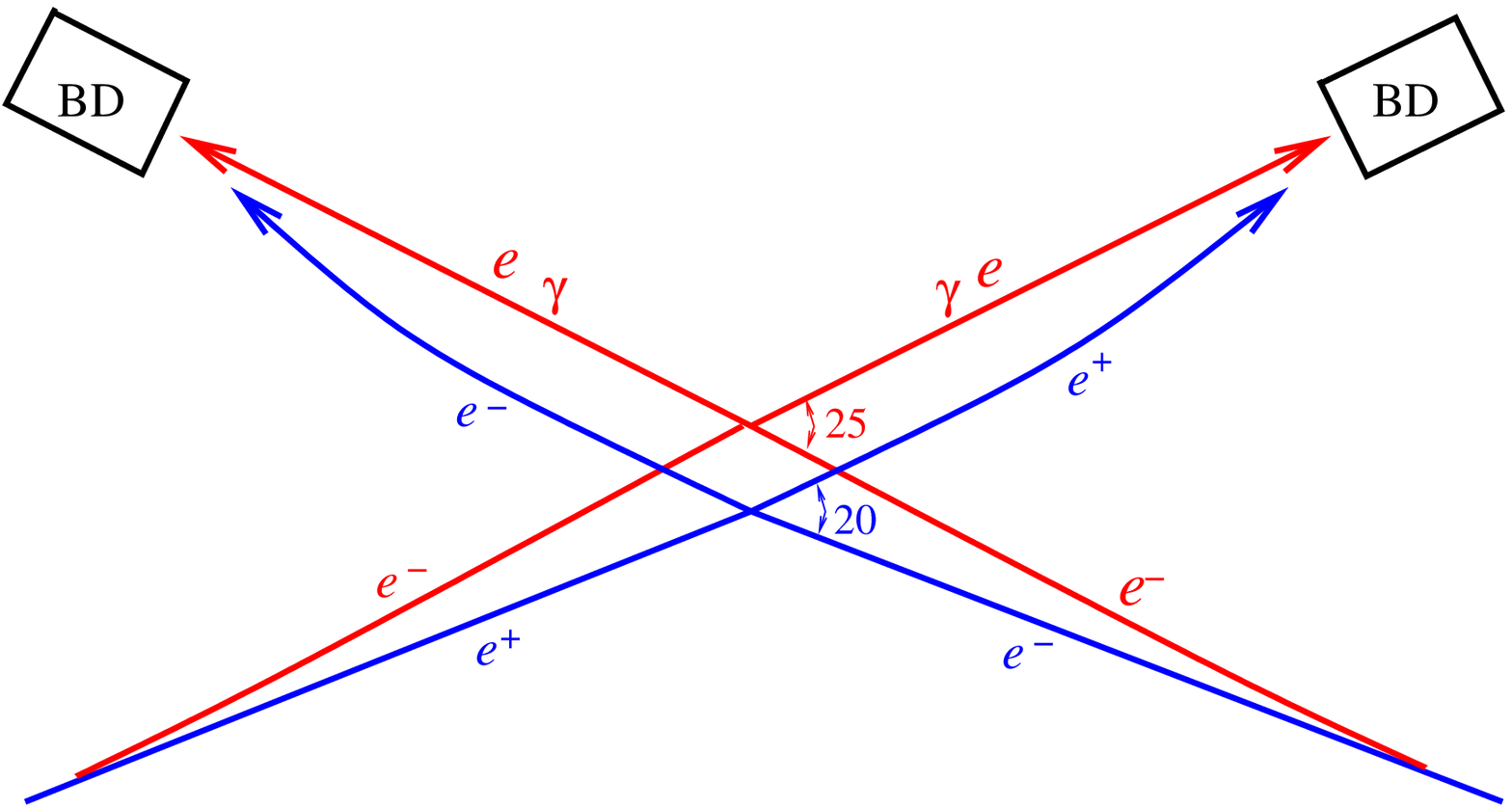}
%\vspace{-1.6cm}
\caption{Possible configurations of the interaction region. The
crossing angle for \EPEM\ may be even somewhat smaller than 20
mrad. } 
\label{config}
\end{figure}

       So, the crab-crossing angle needed for the photon collider
is about 25 mrad, which is, in principle,  compatible with \EPEM.
For \EPEM\ experiments, a smaller angle is preferable, so the angles 2, 14,
20 mrad are considered. Additional considerations important for
the IR design choice are the following: a) it is desirable to have
the same beam dump for \EPEM\ and \GG\ modes (for the same IR), b)
the beam line from the detector to the beam dump for \EPEM\ should
be curved in order to reduce background from the beam dump to the
detector. Three possible configurations are shown in
Fig.~\ref{config}:
\begin{enumerate}
\item In the first scheme, the crossing angle is 25 mrad both for
\GG\ and \EPEM. Only the pathways to the beam dumps  are different for
\EPEM\ and \GG.
\item In the second configuration, the crossing
angle is $25$ mrad for \GG\ and $20$ mrad  (or even somewhat less)
for \EPEM. The detector and beam dump are in the same place for
all modes of collisions, but the beamlines upstream of the detector
(the final focus system) are at different places.
\item In the third scheme,
the crossing angle is $25$ mrad for \GG\ and $20$ mrad  (or even
somewhat less) for \EPEM. Transition from \EPEM\ to \GG\ operation
needs the displacement of the detector and a shift of the final-focus system.
\end{enumerate}

The first scheme is the easiest and needs no displacements, but the
crossing angle may be somewhat larger than necessary for \EPEM.  The
schemes 2 and 3 allow angles smaller than 25 mrad for \EPEM\ 
operation. Though the third scheme needs the displacement of the
detector of about $\mathcal O$(1--3) m (depends on the length of the
straight chromatic correction section at the end of the final focus
system; here it is assumed to be equal to about 500 m
\cite{Raimondi}), it looks more attractive than the second scheme
because it needs smaller total bending angles of beamlines upstream of
the detector and therefore requires a smaller bending length (which is
determined by the emittance dilution; for the  fixed dilution of the
normalize emitance $L \propto E^{3/2}\alpha_b^{5/4}$ ).  The curved
\EPEM\ beamline to beam dump is also easier in the third scheme due to
smaller total bending angle.

So, the third scheme can be recommended for the baseline configuration
and needs optimization by the beam delivery working group (WG4).  It
seems OK for 25 mrad for \GG\ and 20 mrad for \EPEM. The decrease of
the \EPEM\ crossing angle down to 14 mrad is more problematic,
transition from 14 to 25 mrad, certainly, requires more longitudinal
and transverse space.  

At the moment WG4 plans to bend the beams to the interaction points at
the distance about 2000 m from the IP. This length includes also the
collimation system. It makes a problem even for transition from 20$\to$25 mrad
because too wide tunnel is needed. Though it is not clear for me why
the 2.5 mrad bend can not be done after the collimation system at the
distance about 500 m from the IP (as it was assumed above).

\subsection{The {\boldmath \GG\ } luminosity}
The \GG\ luminosity for ILC energies is determined by the geometric
luminosity of electron beams~\cite{TEL_ph99,TEL_NIM_2001,TESLATDR}.
As has already been  mentioned  in the Introduction,  for the nominal ILC beam
parameters the expected \GG\ luminosity in the high-energy peak of the
luminosity spectrum $\LGG\sim 0.17\LEPEM$. However, there is hope to
increase it by a factor of 2--3. There is a general rule: $\LGG \sim
0.1 L_{ \rm geom}$.

So, one needs the smallest beam emittances and beta-functions at the IP,
approaching  the bunch length. Compared to the \EPEM\ case, where
the minimum transverse beam sizes are determined by beamstrahlung and beam
instability, the photon collider needs a smaller product of horizontal
and vertical emittances and smaller horizontal beta-function.
However, there are some problems.

\begin{figure}[!htb]
%\vspace{-0.6cm}
\hspace{0.3cm} \includegraphics[width=17cm]{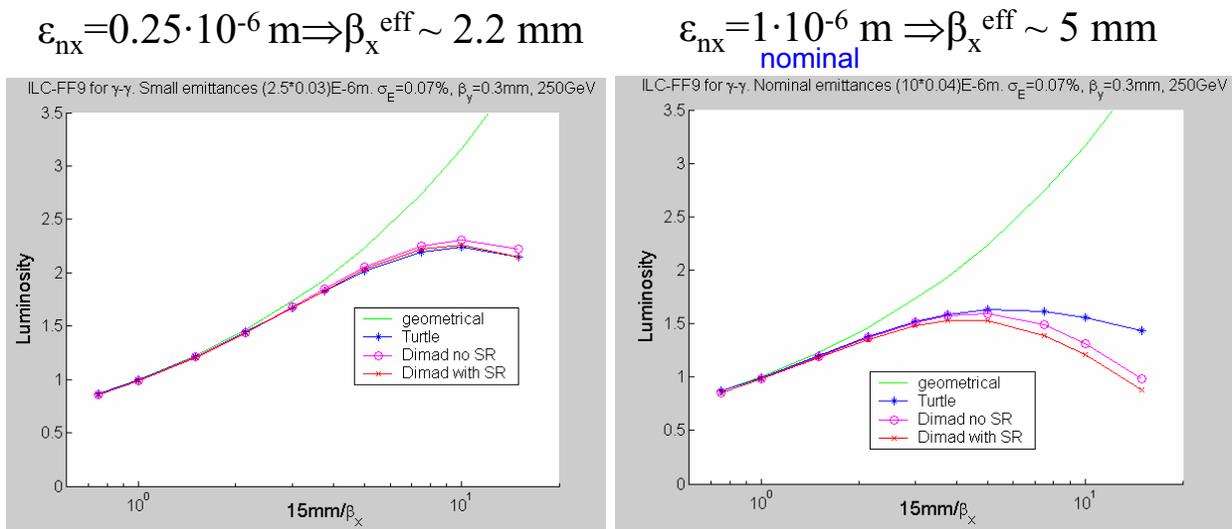}
%\vspace{-1.6cm}
\caption{Dependence of the \GG\ luminosity on the horizontal
  $\beta$-function~\cite{Seryi_snow_gg6}}
\label{seryi}
\end{figure}

The existing final-focus scheme has chromo-geometric aberrations that
limit the effective horizontal beta-function at about 5 mm (for the
nominal horizontal emittance), Fig.~\ref{seryi}~\cite{Seryi_snow_gg6}.
For a horizontal emittance four times lower, the minimum $\beta_x$ is
2.2 mm, which  increases  the \GG\ luminosity by a factor of
3.  So, lower horizontal emittance is very desirable for two reasons.
Present minimum emittances in damping rings originate from
requirements for \EPEM\ collisions, but not from physics limitations
in the damping rings (DR). If we reduce the horizontal emittance by a
factor of 2 and the vertical emittance by 30\%, $\beta_x$ can be
reduced from 5 to 3.7 mm which results in an increase of the \GG\ 
luminosity by a factor of 2. A decrease of the horizontal emittance by
a factor of 4 and by 30\%  the vertical one allows an increase of the
luminosity by a factor of 3.5!  In this case, the high-energy \GG\ 
luminosity will reach almost 60\% of \EPEM\ luminosity, and the number of
events will be greater by about  a factor of 5. This certainly is  a very good
goal.

From A.~Wolski's talk at Snowmass~\cite{Wolski_gg6} it follows that, in
principle, such a decrease of emittances in DR is possible by adding
wigglers in order to reduce the damping time and thus to suppress
intra-beam scattering. It is necessary to study these possibilities in
detail.  Clearly, such a reduction of emittances will increase the DR
cost, but by how much? Such a reduction of emittances would be useful for
\EPEM\ as well, but for \GG\ it means a considerable increase of the
luminosity (which is  time and money).

\subsection{Beam dump}

\begin{figure}[!htb]
\vspace{-1.cm}
 \hspace{0.6cm} \includegraphics[width=7.5cm]{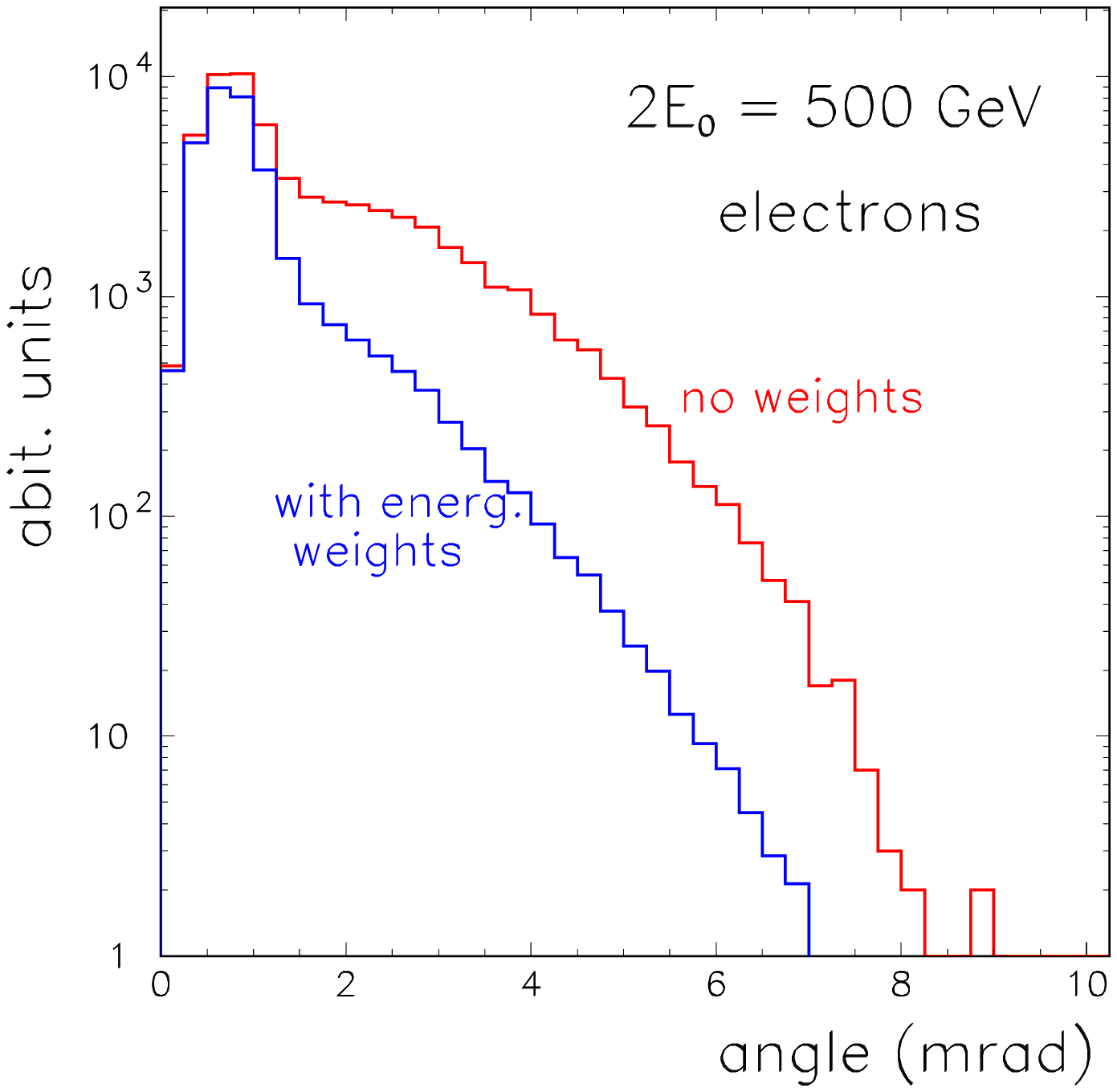}
\hspace{-0.cm}\includegraphics[width=7.5cm]{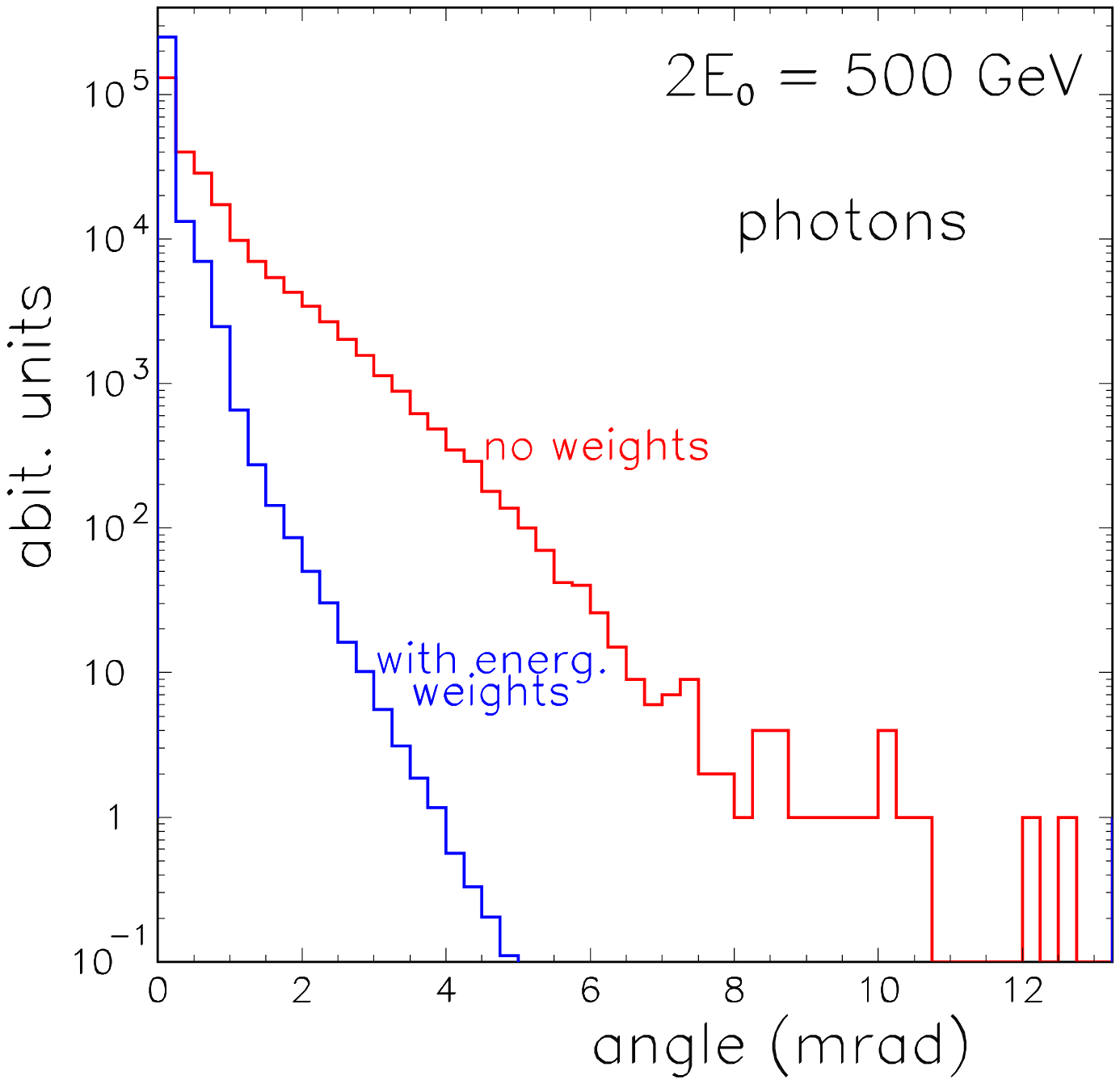}
\vspace{-0.6cm}
\caption{Angular distributions of electrons (left) and photons (right)
after the conversion and interaction points.}
\label{ang-dis}
\end{figure}

\begin{figure}[!htb]
%\vspace{-0.6cm}
 \hspace{0.3cm} \includegraphics[width=7.5cm]{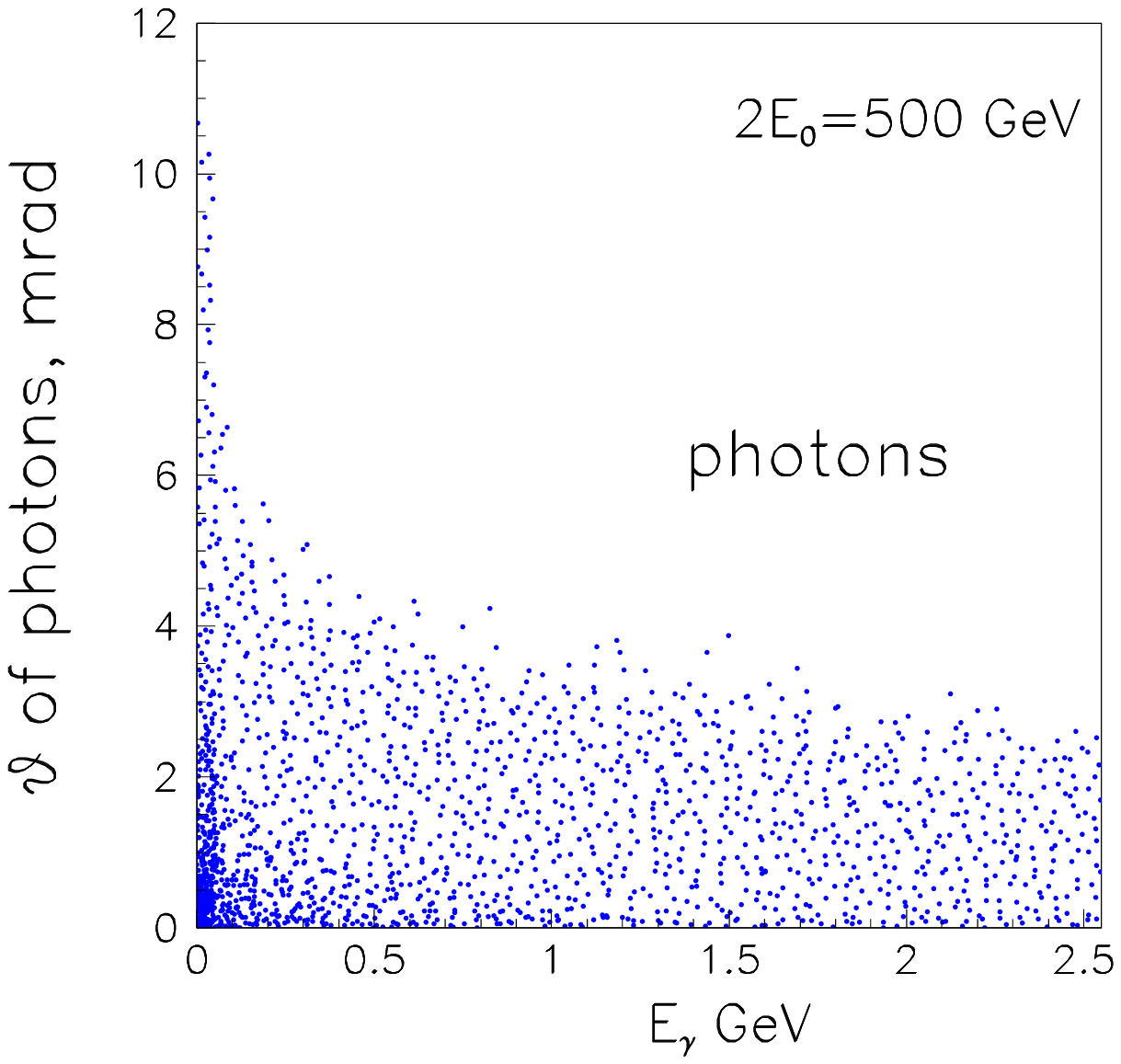}
\hspace{-0.6cm}\includegraphics[width=7.5cm]{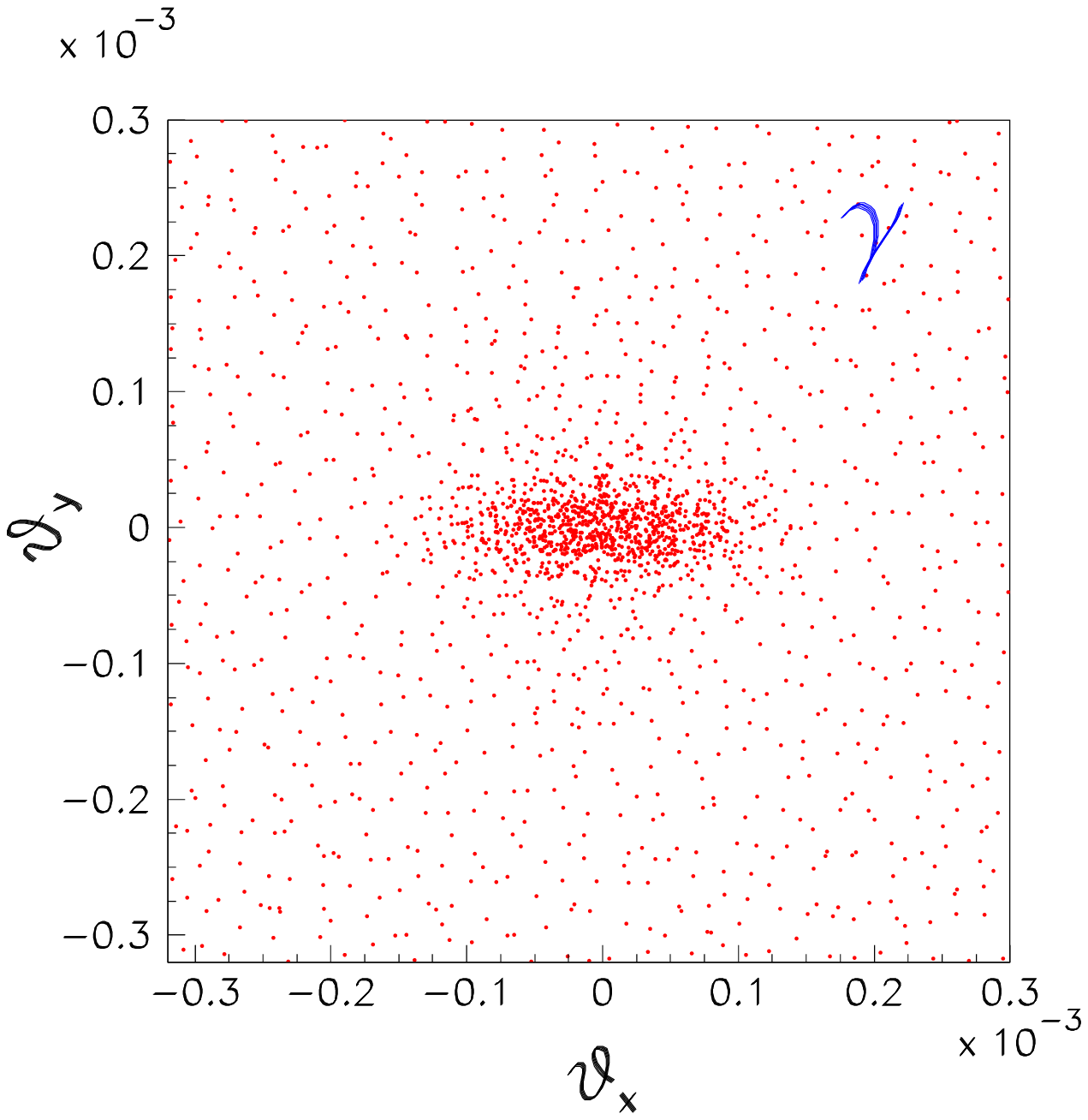}
\vspace{-0.7cm}
\caption{Energy-angular distributions of beamstrahlung photons (left) and
angular distribution of Compton photons (right).}
\label{ang-p}
\end{figure}

\begin{figure}[!htb]
%\vspace{-0.6cm}
\includegraphics[width=16cm]{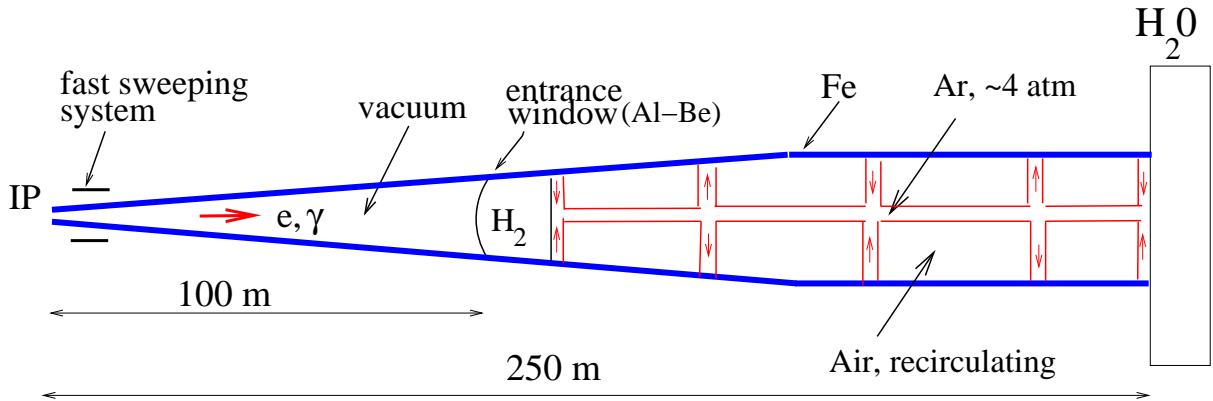}
\vspace{.1cm}
\caption{An idea for the photon collider beam dump.  }
\label{beam-d}
\end{figure}

The photon collider needs a special beam dump, very different from
\EPEM. There are two main differences:
\bi
\item  Disrupted beams at the photon collider is an equal mixture of
  electrons and photons (and some admixture of positrons);
\item Disrupted beams at the photon collider are very wide (see 
  Fig.~\ref{ang-dis}), and need exit pipes with a large diameter.
\item On the other hand, the photon beam after the Compton scattering
  is very narrow. At the distance of 250 m from the IP, the r.m.s.
  transverse size of the photon beam is $1\times 0.35$ mm$^2$, see (see
  Fig.~\ref{ang-p}), with the power of about 10 MW. It cannot be dumped
  directly in a solid or liquid material. \ei
    There is  an idea of such a beam dump and corresponding
  simulations~\cite{Telnov-lcws04}, but the next step required a  more
  careful study. The idea is the following. The water beam dump is
  situated at the distance about 250 m from IP, Fig.~\ref{beam-d}.
The electron beam can be swept by the fast magnets (as in the TESLA
TDR) and its density at the beam dump will be acceptable. In order to
spread the photon beam we suggest to put a gas target, for example Ar
at $P\sim 4$ atm, somewhere between the distance of 120 and 250 m:
photons would produce showers, the beam diameter would increase and
the density at the beam dump would become acceptable.  In order to
decrease the neutron flux in the detector one can add a volume with a
gas hydrogen gas just before the Ar target, which would reduce the flux of
backward-scattered neutrons at the IP at least by one order of
magnitude. The corresponding numbers can be found in
Ref.~\cite{Telnov-lcws04}.

In order to reduce the diameter of the beamline between the beam dump
and the IP, it is desirable to focus somewhat the disrupted electron
(positron) beam just after the exit from the detector (this issue has
not been considered yet). The angular distribution of beamstrahlung photons
is similar to that of beamstrahlung electrons that produced these
photons. However, the energy of beamstrahlung photons produced by
rather  low-energy large-angle electrons is only a small fraction of
their energy, so the
effective (energy-weighted) angular distribution of photons is
narrower than that for electrons. According to Fig.~\ref{ang-dis}
(right), for photons the clear angle $\pm 3$ mrad will be sufficient,
which is 75 cm at the distance of 250 m.

  The Ar target should have the diameter of no more than 10 cm (a shower
  of such diameter  does not present a problem).  The rest
  of the volume of the exit pipe with a diameter of about 1.5 m can be filled
   with air at 1 atm (or vacuum).  Such measures are necessary in
  order to avoid unnecessary scattering of low-energy electrons
  traveling at large distances from the axis and thus to
  reduce the energy losses and activation of materials (water, air) in the
  unshielded area (it is difficult to shield a 100 m tube).

\subsection{Laser system}

 The required laser parameters:
\bi
\item Wavelength \hspace*{1cm} $\sim 1$ \MKM\  (good for $2E_0 < 700$
  GeV); \\[-6mm]
\item Time structure \hspace*{0.8cm} $c \Delta t \sim 100$ m, 3000 bunches/train; \\[-6mm]
\item Flash energy \hspace*{1cm}$\sim 9$ J  (about one scattering length for
  $E_0=250$ GeV); \\[-6mm]
\item Pulse length  \hspace*{1.2cm} $\sigma_t \sim 1.5$ ps.
\ei

The most attractive scheme for a photon collider with the TESLA/ILC pulse
structure is storage and recirculation of a very  powerful laser
bunch in an external optical cavity~\cite{TEL_ph99, TEL_NIM_2001,Will2001,TESLATDR,Klemz2005}. This can reduce
the required laser power by a factor of $\sim$100 (the quality factor of the
cavity).

\begin{figure}[!htb]
\vspace{-1.2cm}
\centering
\includegraphics[width=9cm]{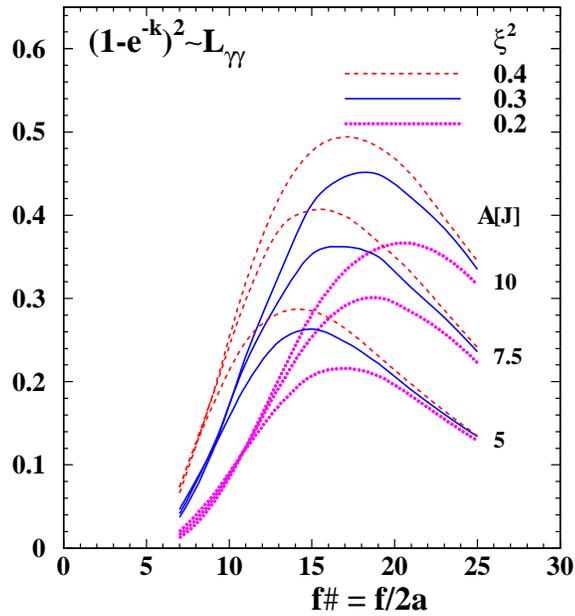}
\vspace{-1.5cm}
\caption{Dependence of  \LGG\ on the flash energy and $f_{\#}$
(flat-top laser beam) for several values of the parameter $\xi^2$. }
\label{conversion}
\end{figure}

Dependence of the \GG\ luminosity on the flash energy and
$f_{\#}=F/2R$ (flat-top laser beam) for several values of the
parameter $\xi^2$ (which characterizes multi-photon effects in Compton
scattering, $\xi^2 < 0.3$ is acceptable \cite{TESLATDR}) is presented
in Fig.~\ref{conversion}~\cite{Telnov_snow05_1}.  This simulation is
based on the formula for the field distribution near the laser focus
for flat-top laser beams. It was assumed that $\alpha_c = 25$ mrad and
the angle between the horizontal plain and the edge of the laser beam
is 17 mrad (the space required for disrupted beams and quads, see
Fig.~\ref{beams-quad}). At the optimum, $f_{\#} \sim 17$, or the
angular size of the laser system is about $\pm 0.5/f_{\#} \approx \pm
30$ mrad.  If the focusing mirror is situated outside the detector at
the distance of 15 m from the IP, it should have a diameter of about 1
m.  All other mirrors in the ring cavity can have smaller diameters,
about 20 cm is sufficient from the damage point of view (diffraction
losses require an additional check).

\begin{figure}[!htb]
%\vspace{-0.6cm}
\includegraphics[width=13cm]{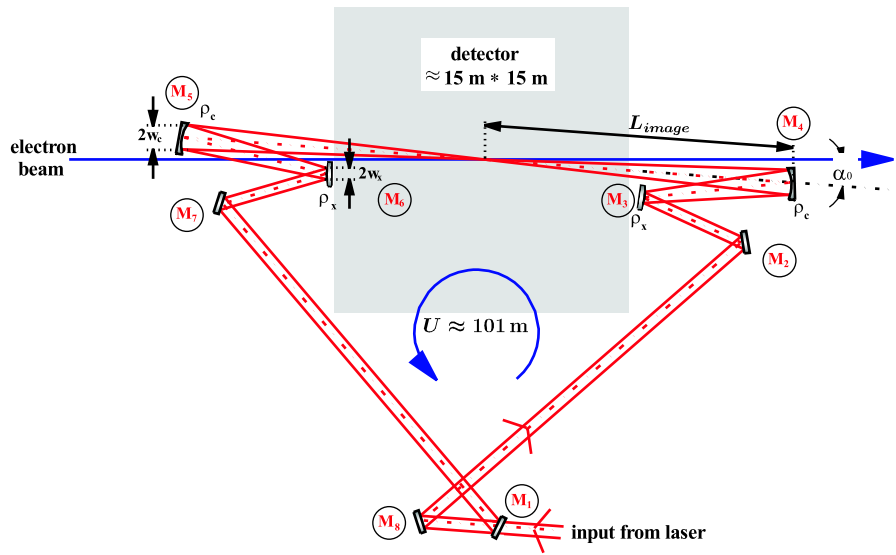}
%\vspace{-1.6cm}
\caption{The ring optical cavity considered in Ref.~\cite{Klemz2005}.}
\label{Klemz1}
\end{figure}

\begin{figure}[!htb]
%\vspace{-0.6cm}
\hspace{1.5cm}\includegraphics[width=12cm]{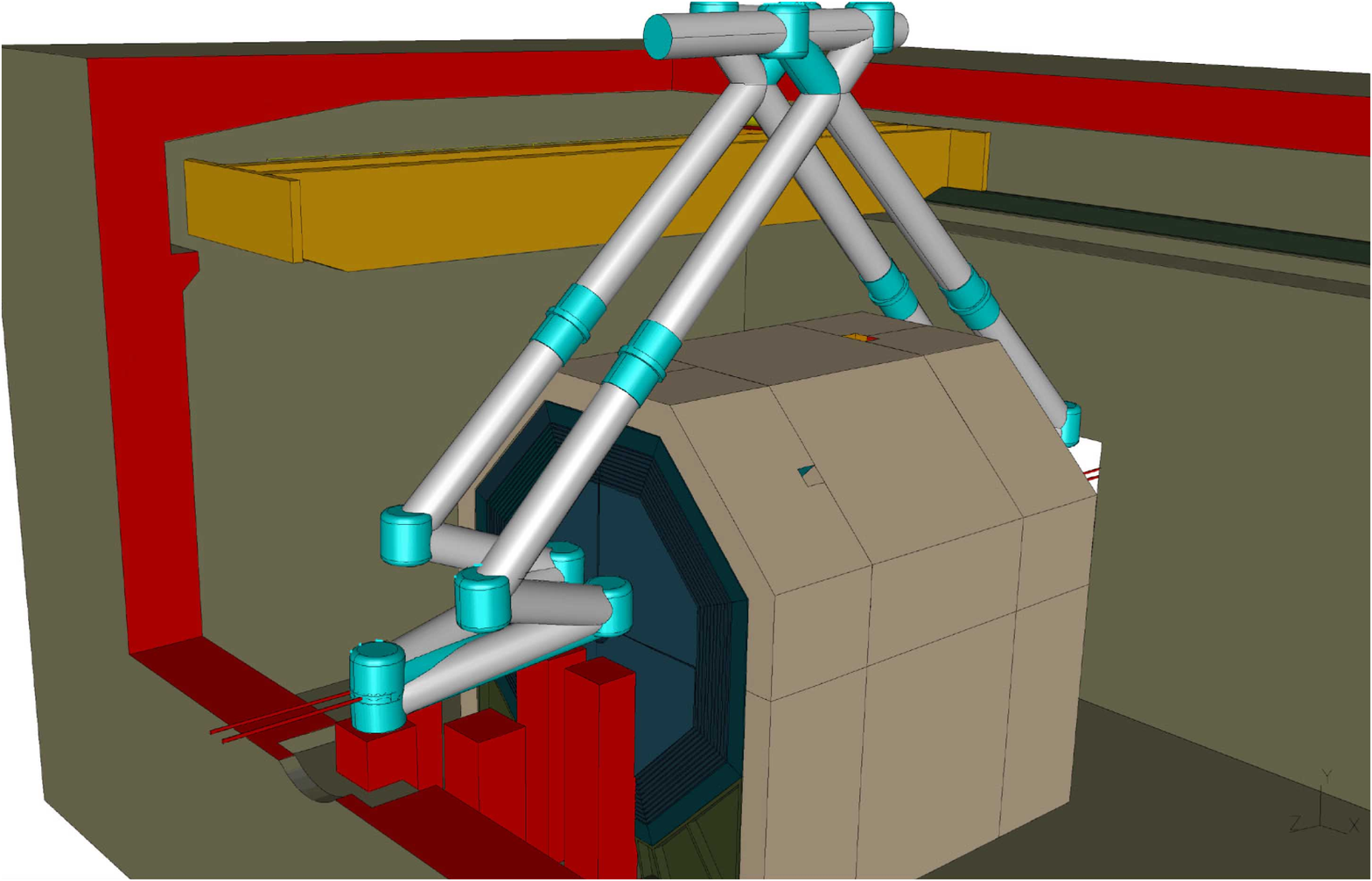}
%\vspace{-1.6cm}
\caption{Layout of the cavity in the detector hall~\cite{Klemz2005}.}
\label{Klemz2}
\end{figure}

The DESY-Zeuthen group has  considered an optical cavity
at the wave level, its pumping by short laser pulses, diffraction
losses, etc.~\cite{Klemz2005}. The scheme of their cavity is shown in
Fig.~\ref{Klemz1}. In this study, truncated Gaussian beams were
used. The results are in agreement with those given above for 
flat-top laser beams. Of course, consideration at the wave level is
much more informative as it allows  simulation of  diffraction losses and
propagation of different laser modes in the cavity. A possible layout of
the cavity in the detector hall is shown in Fig.~\ref{Klemz2}. In this
design, the laser is situated on the surface; it may be   better to
hide the cavity under the detector.


This study of the cavity is only the first step.  Now we need a more
detailed considerations of all technical aspects of such a cavity and
pumping laser with participation of people who are experienced in this field
and are experts in laser technologies. As the result of the next step, we
should understand what is well established and does not need an
additional proofe and what needs additional experimental checks.
Participation of recognized laser experts in these developments and
their critical expertise will be sufficient for convincing the ILC
community and politicians in the feasibility of the photon collider,
but before construction of the real laser system we certainly need to
organize  the experimental laser group and construct a 
prototype in order to get experience and make the final design more
reliable.  There are some plans of such a facility at ATF2 at
KEK~\cite{ATF2}.  Similar, but less powerful, laser cavities are
also developed  for beam diagnostic at the ILC and for the laser
positron source.

Though the cavity reduces substantially the required laser energy, the
laser should still be very powerful. All technologies necessary for
such a laser exist, namely: the chirped pulse technique, adaptive
optics, diode pumping, etc.  According to LLNL estimation, the cost of
one such laser is about \$ 10 M \cite{Gronberg-snow05}.  The photon
collider needs two such lasers and one--two spares.

\begin{figure}[!htb]
\vspace{-2.8cm}
\hspace{-1cm}\includegraphics[width=17cm]{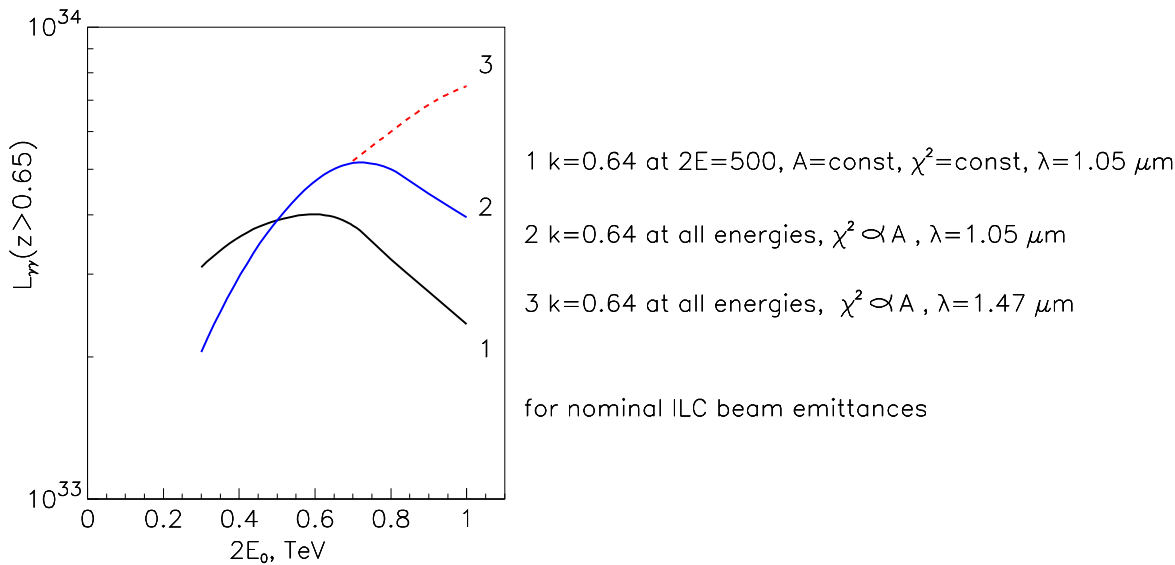}
\vspace{-2.1cm}
\caption{Dependence of the \GG\ luminosity on the energy.}
\label{lgg}
\end{figure}

The same laser with the 1 \MKM\ wavelength can be used up to the ILC
energy $2E_0 \sim 700$ GeV. At higher energies, the \GG\ luminosity
decreases due to \EPEM\ pair creation in the conversion region in
collision of the high-energy and laser photons~\cite{GKST83,TEL95} and
due to the decrease of the Compton cross section, see Fig.~\ref{lgg}.
For the energy $2E_0=1$ TeV, the reduction of the luminosity due to
this effect is about a factor of 2--3 compared to the optimum case.
For the high energies it is desirable to have a wavelength of about
1.5--2 \MKM. The technical feasibility of such a laser has not been
studied yet.

\subsection{Summary on the photon collider. The next steps.}

A crab crossing angle of 25 mrad is the minimum crossing angle
compatible with the photon collider. It is acceptable for \EPEM\ 
operation as well, since the decrease of \LEPEM\ is small.  However,
the \EPEM\ people prefer smaller angles in order to somewhat increase
the detection efficiency for several processes (detection of small
angles is needed for suppresion of some background processes). If the
ILC is optimized only for \EPEM, then the photon collider is not
possible at all. On the other hand, the angle of 20 mrad is considered
as one of possibilities for \EPEM, which is very close to 25 mrad. In
order to facilitate a transition to the photon collider the best would
be the choice of the same angle for \EPEM\ and \GG,\GE, i.e. 25 mrad.
As alternative one can use the scheme 3 in Fig.~\ref{config} which
requires  displacement of the detector and the final focus system. 

In order to fix the crossing angle a  optimize  $L^*$, a  more detailed
simulation of beam losses is required, as well as a more detailed design
of the quad (which already seems  good).

In order to increase \LGG\ it is desirable to decrease emittances in
the DRs.

  It is necessary to develop  the final focus
system for the photon collider with a small $\beta_x$ and understand
whether it is compatible with \EPEM\  or needs different hardware.

 There exist  ideas  for the photon collider beam dump, detailed
consideration is necessary.

There are some studies of the laser optical cavity for the
photon collider, the next steps is  consideration of all
technological aspects, which requires participation of laser experts (needs
money).

   At the photon collider, the angle of about $\pm 100$ mrad is occupied by
laser beams; this  should be taken into account in the design of one
of the detectors.

A decision about the status of the photon collider in ILC project
should be made expediciuosly, because a) the photon collider
determines design of many ILC elements; b) people will join to the
development of the photon collider only if it is a part of the ILC and
has political and financial support.

\section{ELECTRON-ELECTRON COLLIDER}
Electron-electron collider presents a very unique possibility for
study of many phenomena at the ILC under very clean conditions
(without background from annihilation processes).  Physics in \EMEM\ 
collisions was discussed at several \EMEM\ workshops (organized by
C~.Heusch) and proceedings are published~\cite{emem}.  Such type of
collisions requires minimum modification of the ILC, mainly in the
final-focus system;  nevertheless, this requires attention of accelerator
people.  Due to the beam repulsion, the attainable luminosity is by a
factor of 5 lower than in \EPEM\ collisions.  At Snowmass,
P.Bambade~\cite{Bambade-snow05} discussed the possibility of \EMEM\ in
the scheme with a 2 mrad collision angle (where quads deflect outgoing
beams). It was shown that the \EPEM\ final focus system can be
readjusted to \EMEM\ in the case of more-round than optimal beams,
with an additional loss in the luminosity by a factor of 2 and larger
beamstrahlung. More studies are needed.

In summary: this option is important, and though seems simple
technically (change of the magnet polarity),  in reality its
realization needs careful consideration of all accelerator parts, and
solutions are not always simple.

\section{GIGA-Z}
Physics requirements for GigaZ were considered at Snowmass by
K.~M\"{o}nig~\cite{Monig-GZ}. Only one remark here. GigaZ needs
polarized beams with a small energy spread. The scheme with an
undulator at $E \approx 150$ GeV and further deceleration gives a large
energy spread, so a bypass is needed. Effects of the wakefield
in the main linac in GigaZ operation was considered at Snowmass by
K.~Kubo~\cite{Kubo}. Undulator positron source in the GigaZ option was
discussed by D.~Scott \cite{Scott}.

\section{FIXED TARGET EXPERIMENTS}
\begin{figure}[!htb]
%\vspace{-2.8cm}
\hspace{0cm}\includegraphics[width=15cm]{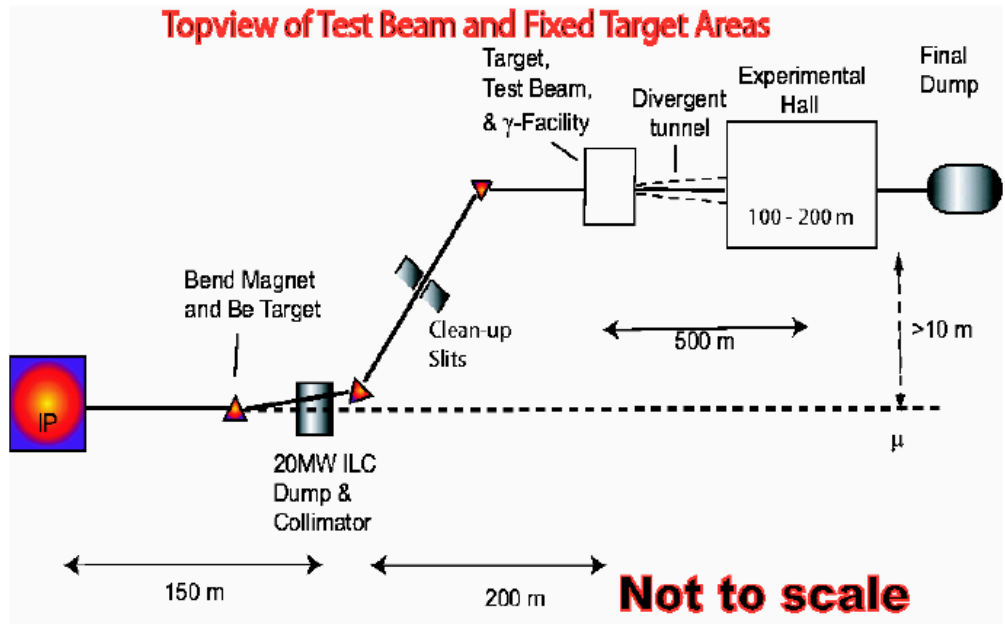}
%\vspace{-2.6cm}
\caption{Fixed target area at the ILC.}
\label{mtingwa}
\end{figure}

Fixed target experiments are traditional in  particle physics and
should be not ignored at the ILC~\cite{Mtingwa, Kolomensky, Kanemura}.
For this purpose one can use the spent electron beam after the IP and
deflect it from the beam dump to the experimental area. It consists
of clean-up slits, a target area for production of the photon beam using
crystal or laser targets, divergent tunnel, experimental hall and final
dump. All together occupy more than one km in length,
Fig.~\ref{mtingwa}~\cite{Mtingwa}.  In order to reduce overlapping of
events, it is attractive to consider the possibility of filling the ILC
train by low-charge bunches between the primary ILC bunches and to deflect
them by kickers for fixed target experiments.

If this option is accepted, it considerably influences the design of the
ILC interaction region.


\begin{thebibliography}{99}
\bibitem{TESLATDR} B.~Badelek et. al., The Photon Collider at TESLA,
{\it Intern. Journ. Mod. Phys.} {\bf A 30} (2004) 5097-5186, hep-ex/0108012.

\bibitem{Telnov_snow05_1} V.~I.~Telnov, Photon collider at ILC, talk at
  the Second ILC Accelerator workshop, Snowmass, GG6 group, Colorado, August 14-27,2005.

\bibitem{Parker} B.~Parker, QDO external field compensation
  possibilities for gamma-gamma, talk at the Second ILC Accelerator
  workshop, Snowmass, GG6 group, Colorado, August 14-27,2005.

\bibitem{telnov_lcws05_1113} V.~I.~Telnov, Crossing angle at the photon
 collider, the talk at Intern. Linear Collider Workshop (LCWS 2005),
 Stanford, California, 18-22 Mar 2005,  physics/0507134.



\bibitem{TEL_ph99} V.~I.~Telnov,  Status of gamma gamma, gamma
  electron colliders, Proc. of Intern. Conf. on the Structure and
  Interactions of the Photon (Photon 99), Freiburg, Germany, 23-27 May
  1999, published in  Nucl.\ Phys.\ Proc.\ Suppl.\  {\bf 82} (2000) 359.

\bibitem{TEL_NIM_2001} V.~I.~Telnov, Photon collider at TESLA, Nucl.\ Instrum.\ Meth.\ A {\bf
  472} (2001) 43, hep-ex/0010033.

\bibitem{Seryi_snow_gg6} A.~Seryi, Discussion of gamma-gamma
  parameters, talk at the Second ILC Accelerator workshop, GG6 group, Snowmass,
  Colorado, August 14-27,2005.

\bibitem{Wolski_gg6} A.~Wolski,  Low-Emittance Issues for ILC Damping
  Rings, talk at the Second ILC Accelerator workshop, GG6 group, Snowmass,
  Colorado, August 14-27,2005.

\bibitem{Raimondi}
  P.~Raimondi and A.~Seryi,  Phys.\ Rev.\ Lett.\  {\bf 86}, 3779 (2001).

\bibitem{Telnov-lcws04} L.~I.~Shekhtman and V.~I.~Telnov, A conception
  of the photon collider beam dump, physics/0411253. Proc. of
  Intern. Conf. on Linear Colliders (LCWS 04), Paris, France, 19-24
  Apr 2004.

\bibitem{Will2001} I.~Will, T.~Quast, H.~Redlin and W.~Sandner,
  A laser system for the TESLA photon collider based on an external ring
  resonator,  Nucl.\ Instrum.\ Meth.\ A {\bf 472} (2001) 79.

\bibitem{Klemz2005}  G.~Klemz, K.~Monig and I.~Will,
  Design study of an optical cavity for a future photon collider at
  ILC, DESY-05-098,  physics/0507078.

\bibitem{ATF2} B.~I.~Grishanov {\it et al.}  [ATF2 Collaboration],
 ATF2 proposal, SLAC-R-771.

\bibitem{Gronberg-snow05} G.~Gronberg,  Options Photon Collider Laser
  Facilities,  talk at the Second ILC Accelerator workshop, GG6 group, Snowmass,
  Colorado, August 14-27,2005.

\bibitem{emem} Proceedings of the electron-electron linear collider
  workshops, ed. C.Heusch, Intern. J. of Mod. Physics A 13 (1995) 2217,
  15 (2000) 2347, 18 (2003) 2733.

\bibitem{Bambade-snow05} P.~Bambade, Which beam parameters for
\EMEM?
  Talk at the Second ILC Accelerator workshop, GG6 group, Snowmass,
  Colorado, August 14-27,2005.


\bibitem{Monig-GZ} K.~M\"{o}nig, Physics requirements for GigaZ, Talk at the
  Second ILC Accelerator workshop, GG6 group, Snowmass, Colorado,
  August 14-27,2005.

\bibitem{Kubo} K.Kubo,  Effect of Wake in Main Linac in Giga-Z Operation, Talk at the
  Second ILC Accelerator workshop, GG6 group, Snowmass, Colorado,  August 14-27,2005.

\bibitem{Scott} D.~Scott,  Undulator positron source and GigaZ option,  Talk at the
  Second ILC Accelerator workshop, GG6 group, Snowmass, Colorado,
  August 14-27,2005.

\bibitem{Mtingwa}  S.~Mtingwa, Fixed target experiments at ILC: overview, Second
 ILC Accelerator workshop,
 GG6 group, Snowmass, Colorado,
  August 14-27,2005.

\bibitem{Kolomensky}  Yu.~Kolomensky, General Perspectives and M\"{o}ller Scattering, Second
 ILC Accelerator workshop, GG6 group, Snowmass, Colorado,  August
 14-27,2005.
\bibitem{Kanemura} S.~Kanemura, A Possibility of Measuring LFV
  couplings through the Deep Inelastic Process of eN $\to \tau$ X, Second
  ILC Accelerator workshop, GG6 group, Snowmass, Colorado, August
  14-27,2005.

\bibitem{GKST83} I.~F.~Ginzburg, G.~L.~Kotkin, V.~G.~Serbo, V.~I.~Telnov, {\it
Nucl. Instr.  {\rm\&} Meth.} {\bf 205} (1983) 47.

\bibitem{TEL95} V.~I.~Telnov, {\it Nucl. Instr.  {\rm\&} Meth.} {\bf
355} (1995) 3.

\end{thebibliography}
\end{document}